\documentclass[conference]{IEEEtran}
\IEEEoverridecommandlockouts
\usepackage{cite}
\usepackage{amsmath,amssymb,amsfonts}
\usepackage{algorithmic}
\usepackage{graphicx}
\usepackage{textcomp}
\usepackage{xcolor}
\usepackage{titlesec}
\usepackage{tabularx}
\usepackage{setspace}
\usepackage{caption}
\usepackage{float}
\usepackage{subcaption}
\def\BibTeX{{\rm B\kern-.05em{\sc i\kern-.025em b}\kern-.08em
		T\kern-.1667em\lower.7ex\hbox{E}\kern-.125emX}}
\begin{document}
	%    \begin{spacing}{0.93}
		\title{A Comprehensive Study of PAPR Reduction Techniques for Deep Joint Source Channel Coding in OFDM Systems}
		
		\author{\IEEEauthorblockN{Maolin Liu\textsuperscript{1,2}, Wei Chen\textsuperscript{1,2,3}, Jialong Xu\textsuperscript{1,2}, Bo Ai\textsuperscript{1,2,4}}
			\IEEEauthorblockA{\textsuperscript{1}State Key Laboratory of Advanced Rail Autonomous Operation, Beijing Jiaotong University, China\\
			\textsuperscript{2}School of Electronic and Information Engineering, Beijing Jiaotong University, Beijing, China\\
		    \textsuperscript{3}Key Laboratory of Railway Industry of Broadband Mobile Information Communications\\
	        \textsuperscript{4}Beijing Engineering Research Center of High-speed Railway Broadband Mobile Communications\\
	       % \textsuperscript{5}School of Information Engineering, Zhengzhou University, Zhengzhou, China\\
           Corresponding author: \emph{Wei Chen}
}
%			\thanks{This work is supported by the Natural Science Foundation of China (62122012, 62221001); the Beijing Natural Science Foundation (L202019, L211012); the Fundamental Research Funds for the Central Universities (2022JBQY004).}
		}
		
\maketitle
\begin{abstract}
Recently, deep joint source channel coding (DJSCC) techniques have been extensively studied and have shown significant performance with limited bandwidth and low signal to noise ratio. Most DJSCC work considers discrete-time analog transmission, while combining it with orthogonal frequency division multiplexing (OFDM) creates serious high peak-to-average power ratio (PAPR) problem. This paper conducts a comprehensive analysis on the use of various OFDM PAPR reduction techniques in the DJSCC system, including both conventional techniques such as clipping, companding, SLM and PTS, and deep learning-based PAPR reduction techniques such as PAPR loss and clipping with retraining. Our investigation shows that although conventional PAPR reduction techniques can be applied to DJSCC, their performance in DJSCC is different from the conventional split source channel coding. Moreover, we observe that for signal distortion PAPR reduction techniques, clipping with retraining achieves the best performance in terms of both PAPR reduction and recovery accuracy. It is also noticed that signal non-distortion PAPR reduction techniques can successfully reduce the PAPR in DJSCC without compromise to signal reconstruction.
\end{abstract}
		
		
\section{Introduction}
%        Conventional digital communications based on the separation theorem \cite{ref2} may no longer be able to meet the potential requirements of next-generation wireless communication systems (B5G/6G), such as emerging wireless applications for brain-computer interaction, autonomous driving, smart manufacturing, and telemedicine, where the requirements for low latency and high reliability are further enhanced.
% Joint source channel coding (JSCC) is a technique used in wireless communication systems that combines the encoding of source information with the modulation of the signal sent over the communication channel. This integration effectively reduces the overall transmission error rate and decreases the amount of bandwidth needed for signal transmission \cite{68,5563107}. JSCC helps optimize the use of available resources, making wireless communication more efficient and cost-effective. Additionally, JSCC can improve the perceived quality of wireless communication, enabling the creation of more complex data-intensive applications that utilize high-quality audio, video, and other multimedia content. With the increasing demand for high-speed wireless data transmission for applications such as video conferencing and virtual reality, JSCC has become an important technology to enhance the performance of wireless communication systems.
Deep learning (DL) is considered as a promising technique in solving wireless communication problems in recent years \cite{73,74,ref3,ref6}. In particular, the deep joint source channel coding (DJSCC) approach \cite{ref3,ref6} has been demonstrated to be more effective than the traditional split source channel coding (SSCC) technique, with lower distortion, simpler design and computation, and the elimination of the ``cliff effect". However, due to the randomness of the generated symbols, DJSCC poses a serious high peak-to-average power ratio (PAPR) problem when combined with orthogonal frequency division multiplexing (OFDM). High PAPR results in in-band distortion and out-of-band radiation, which are caused by the nonlinearity of the radio transmitter's high-power amplifier (HPA). Moreover, the detection efficiency of OFDM receivers is particularly sensitive to nonlinear devices such as digital-to-analog converters (DACs) and HPA.

\begin{table*}[!t]
%\renewcommand\arraystretch{1}
%\vspace{0.3in}
\centering
\caption{Comparison of the PAPR reduction techniques for DJSCC studied in literature and our paper.}  % 表格标题
\label{Table1}
\begin{tabular}{c|c|c c c c}
		\hline
Category & PAPA reduction technique& \cite{66} & \cite{65} & Ours \\
		\hline
Conventional signal&Clipping & \checkmark & \checkmark & \checkmark \\
distortion technique &Companding &  &  & \checkmark \\
		\hline
DL-based signal&PAPR Loss &  & \checkmark &  \checkmark\\
distortion technique&Clipping with retrain & \checkmark & \checkmark & \checkmark \\
		\hline
Signal&SLM &  &  & \checkmark \\
non-distortion technique&PTS &  &  & \checkmark \\
\hline
\end{tabular}
\end{table*}

Various PAPR reduction techniques have been proposed and successfully used in engineering practice. PAPR reduction techniques can be classified into signal distortion and non-distortion classes depending on whether they introduce distortion to the original signal or not. The signal distortion PAPR reduction techniques mainly include clipping\cite{33}, clipping filtering\cite{34}, peak-adding windowing\cite{36}, peak-canceling\cite{39} and companding\cite{41}. Signal non-distortion PAPR reduction techniques include precoding\cite{43}, selective mapping (SLM)\cite{45}, partial transmit sequence (PTS)\cite{45}, interleaving\cite{69}, tone reservation\cite{47}, tone injection\cite{47}, active constellation expansion\cite{48}, and constellation reshaping\cite{49}. In addition to conventional schemes, some DL-based PAPR reduction techniques have recently been proposed \cite{51}, whose main idea is to model and train the OFDM system end-to-end using a neural network model, leading to low PAPR output.

\begin{figure}[t]
%\vspace{0.2in}
\centering
\includegraphics[width=3.4in]{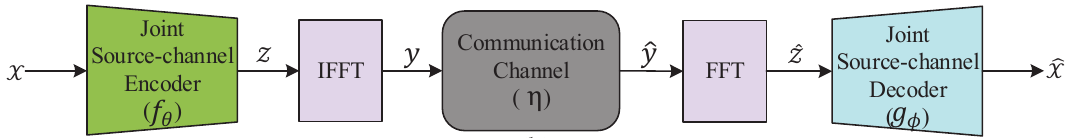}
\caption{The OFDM-based DJSCC system.}
%\vspace {-0.4cm}
\label{fig_1}
\end{figure}

Few investigations have been conducted into the problem of high PAPR in OFDM-based DJSCC systems.
% In \cite{66}, clipping is applied to the DJSCC framework. Furthermore, a PAPR sensitive loss is designed in \cite{65} to trade off the PAPR and the recovery performance.
These works only consider signal distortion PAPR reduction methods, and there is no investigation on the effectiveness of signal non-distortion methods for DJSCC as far as we know. Furthermore, even for signal distortion methods, existing work focuses mainly on the clipping technique\cite{66,65}. In view of the fact that there exist many PAPR reduction methods, no work has comprehensively studied and compared different methods for DJSCC. To this end, in this paper, we evaluate the performance of various PAPR reduction techniques for OFDM in DJSCC systems. The coverage of the PAPR reduction techniques investigated in this paper and the related literature \cite{66,65}  are summarized in Table \ref{Table1}.

\section{System Model}
\subsection{Deep joint source-channel coding}
Considering a point-to-point DJSCC based OFDM system for signal transmission as shown in Fig. \ref{fig_1}, the encoder maps the $n$-dimensional input $\boldsymbol{x} \in \mathbb{R}^n$ to a $k$-dimensional complex-valued feature $\boldsymbol{z} \in \mathbb{C}^k$, and $\boldsymbol{z}$ satisfies the average power constraint $\frac{1}{k}\mathbb{E}(\boldsymbol{z}\boldsymbol{z}^\ast) \leq 1$, where $\boldsymbol{z}^\ast$ represents the complex conjugate of $\boldsymbol{z}$. The bandwidth to signal ratio is defined as $R=\frac{k}{n}$. The encoding function $f_\theta : \mathbb{R}^n \to \mathbb{C}^k$ is a neural network with a set of parameters $\theta$, and the encoding process can be represented as $\boldsymbol{z} = f_\theta(\boldsymbol{x})$. The encoded symbol sequence $\boldsymbol{z}$ is transformed by inverse fast Fourier transform (IFFT) into a time domain channel input $\boldsymbol{y}$ and sent through an additive Gaussian white noise channel (AWGN). The channel output $\hat{\boldsymbol{y}} \in \mathbb{C}^k$ can be expressed as $\hat{\boldsymbol{y}} = \boldsymbol{y} + \boldsymbol{n}$, where $\boldsymbol{n} \in \mathbb{C}^k$ is an independent identically distributed Gaussian noise vector with mean 0 and variance $\sigma^2$.
		
The received signal $\hat{\boldsymbol{y}}$ is transformed by fast Fourier transform (FFT) to generate the frequency domain signal $\hat{\boldsymbol{z}}$, which is used as the input of the decoder. The decoder maps $\hat{\boldsymbol{z}}$ to $\hat{\boldsymbol{x}} \in \mathbb{R}^n$ that is the estimation of the original vector $\boldsymbol{x} \in \mathbb{R}^n$. Similarly to the encoder, the decoder function $g_\phi : \mathbb{C}^k \to \mathbb{R}^n$ exploits a neural network with a set of parameters $\phi$. The decoding process can be expressed as:
			\begin{equation}\label{eq:1-0}
			%			\setlength{\abovedisplayskip}{1pt}
			%			\setlength{\belowdisplayskip}{1pt}
			%			\label{deqn_ex1a}
			\hat{\boldsymbol{x}} = g_\phi(\hat{\boldsymbol{z}}) = g_\phi(FFT(f_\theta(IFFT(\boldsymbol{x})))+\boldsymbol{n}).
		\end{equation}
The distortion between the original signal $\boldsymbol{x}$ and the reconstructed signal $\hat{\boldsymbol{x}}$ can be expressed as $d(\boldsymbol{x},\hat{\boldsymbol{x}}) = \frac{1}{n}\sum_{i=1}^{n} (x_i - \hat{x}_i)^2$,
where $x_i$ and $\hat{x}_i$ represent the $i$th elements of $\boldsymbol{x}$ and $\hat{\boldsymbol{x}}$, respectively. With a fixed bandwidth to signal ratio $R$, the DJSCC design aims to find the encoder and decoder parameters $\theta^*$ and $\phi^*$ that minimize the end-to-end distortion as follows:
		\begin{equation}\label{eq:2}
%				\setlength{\abovedisplayskip}{1pt}
%			\setlength{\belowdisplayskip}{1pt}
%			\label{deqn_ex2a}
			(\theta^*,\phi^*) = \mathop{\arg\min}\limits_{\theta,\phi^*}\mathbb{E}_{p(\boldsymbol{x},\hat{\boldsymbol{x}})}[d(\boldsymbol{x},\hat{\boldsymbol{x}})],
		\end{equation}
where $p(\boldsymbol{x},\hat{\boldsymbol{x}})$ is the joint probability distribution of the original and reconstructed signal. We model the encoder and decoder using the neural network structure in \cite{ref3} for image transmission.
	
%		\begin{figure*}[t]
%			\centering
%			\includegraphics[width=7in]{tranceiver}
%			\caption{Schematic diagram of the SDR-based OFDM communication system for DJSCC. }
%			\label{fig_2}
%			
%			\setlength{\belowcaptionskip}{-1cm}
%		\end{figure*}
		
\subsection{Peak-to-average power ratio in OFDM}
%OFDM is one of the multicarrier modulation (MCM) techniques, which decomposes high-speed serial data streams into multiple independent low-speed data streams and modulates them onto mutually orthogonal subcarriers for parallel transmission, thus being able to counteract the delay expansion caused by multipath effects and greatly improve spectral efficiency.
%		The discrete OFDM baseband signal with carrier number $N$ can be expressed as:
%		\begin{equation}\label{eq:3}
%			x_{k} = \frac{1}{\sqrt{N}}{\sum\limits_{n = 0}^{N - 1}{X_{n}e^{j2\pi kn\frac{\mathrm{\Delta}fT}{L}}}}
%		\end{equation}
%	    where $x_{k}$ represents the time-domain OFDM signal, $k\in 0,1,2,3,\ldots$, $NL - 1$. $X_{n}$ represents the frequency domain modulation symbol, $n\in 0,1,2,3,\ldots,N - 1$, and $L$ represents $L$ times oversampling. It can be seen that the sequence ${x_{k}}$ can be interpreted as a discrete Fourier inverse transform (IDFT) with $(L - 1)N$ zero-filled data blocks $X$, and the IFFT and FFT are commonly used in engineering to implement OFDM modulation and demodulation.
	
The PAPR in OFDM is defined as the ratio of the peak power to the average power of the signal, and the PAPR of a discrete OFDM signal can be expressed as follows:
\begin{equation}\label{eq:4}
PAPR\left( {dB} \right) = 10{\log_{10}\frac{\max\limits_{k}{y_{k}}^{2}}{E\left\lbrack \left\| \boldsymbol{y} \right\|_{2}^{2} \right\rbrack}}
\end{equation}
where the denominator $E\lbrack \cdot \rbrack$ denotes the expectation, $y_{k}$ denotes the $k$-th sample of the time-domain OFDM symbol, $k\in\{0,1,2,... ,N-1\}$. The superposition of multiple carrier signals leads to large peaks in the OFDM signal, and this phenomenon is more severe as the number of subcarriers increases. The performance of PAPR is usually evaluated using the complementary cumulative distribution function (CCDF) of PAPR, i.e., $CCDF\left( \text{PAPR0} \right) = P\left\{ \text{PAPR} > \text{PAPR0}\right\}$.

        % \begin{figure}[t]
        % 	\centering
        % 	\includegraphics[width=1.73in]{fig_2.eps}
        	
        % 	\caption{PAPR performance comparison between DJSCC and 16QAM.}
        % 	\label{fig_2}
        % 	\setlength{\belowcaptionskip}{-5cm}
        % \end{figure}
        % \begin{figure}[t]
        % 	\centering
        % 	\includegraphics[width=1.9in]{djscc_constellation.eps}
        	
        % 	\caption{DJSCC constellation diagram.}
        % 	\label{fig_21}
        % 	\setlength{\belowcaptionskip}{-5cm}
        % \end{figure}
        \begin{figure}[t]
        \centering
        \subcaptionbox{\label{2_a}}{\includegraphics[width=1.6in]{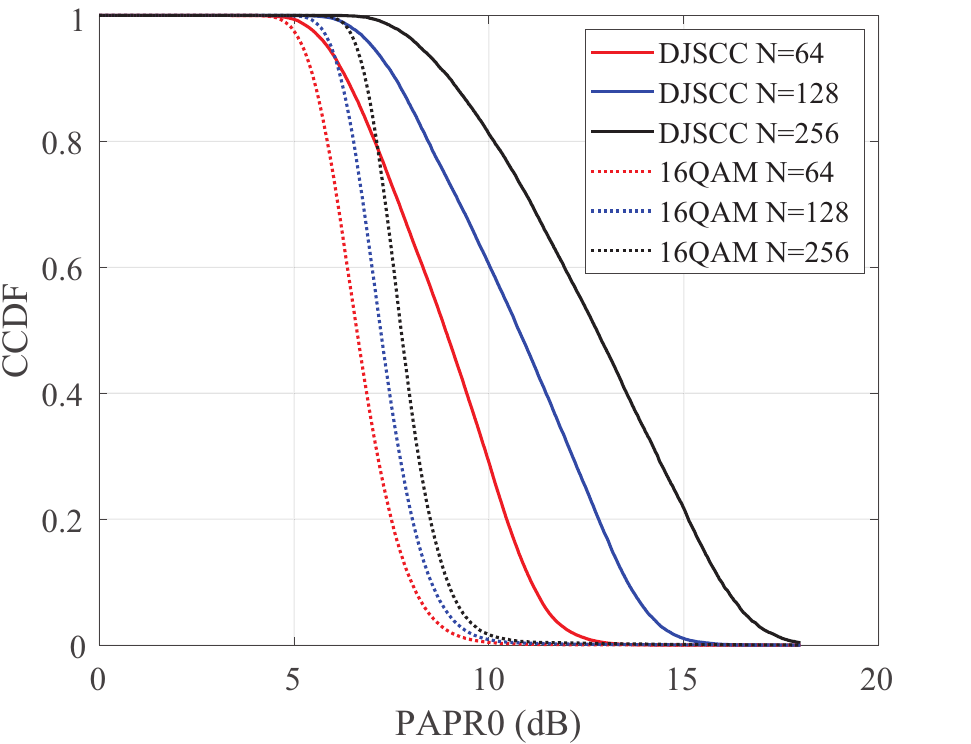}}
        \hfill
        \subcaptionbox{\label{2_b}}{\includegraphics[width=1.65in]{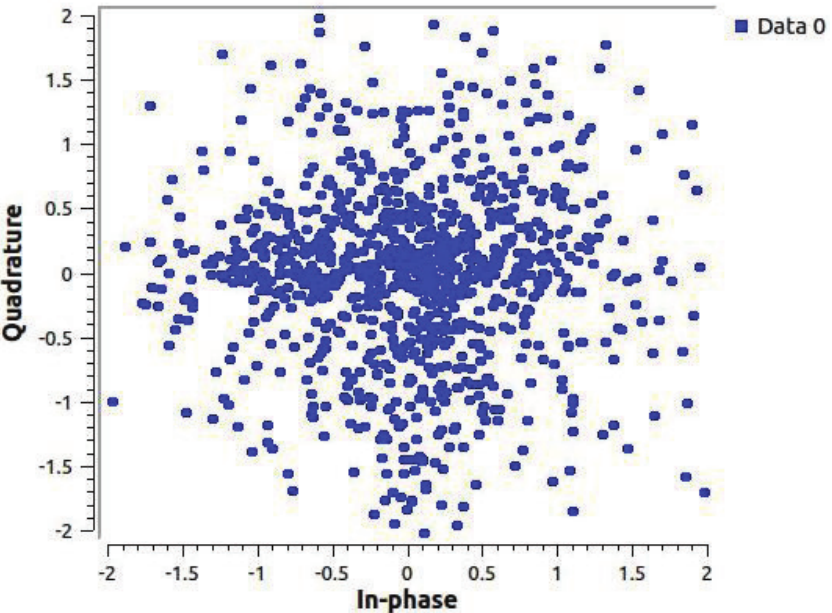}}
        \caption{(a) PAPR performance comparison between DJSCC and 16QAM. (b) DJSCC constellation diagram.}
        \label{fig_2}
        \end{figure}

We evaluate the PAPR performance of DJSCC and 16QAM in the OFDM system with different numbers of subcarriers, and the result is shown in Fig. \ref{2_a}. It can be seen that the PAPR values gradually increase as the number of carriers increases, but DJSCC has a larger PAPR, and the increase in the number of carriers intensifies this phenomenon. This can be explained by the fact that the DJSCC discrete-time analog transmission generates scattered and random modulation constellation points as shown in Fig. \ref{2_b}, which are more likely to generate large peak power symbols, and the superposition of multiple carriers further exacerbates this phenomenon, thus producing a higher PAPR.

\section{PAPR Reduction Techniques}
% High PAPR raises the requirements for ADC/DAC accuracy and power amplifier linear amplification interval in RF hardware devices, increasing hardware costs and even introducing signal nonlinear distortion in severe cases.
In this section, conventional PAPR reduction techniques and deep learning-based PAPR reduction techniques are introduced.
		
\subsection{Conventional signal distortion techniques}
\subsubsection{Clipping}
Clipping simply crops the signal according to a set threshold, reducing the signal peak while retaining most of the low amplitude signal samples, the clipped signal can be expressed as:
		\begin{equation}\label{eq:6}
			y_{clip}(n) = \left\{ \begin{matrix}
				{y\lbrack n\rbrack e^{j\phi{\lbrack n\rbrack}},y\lbrack n\rbrack < \rho\sqrt{\bar{P}}} \\
				{\rho\sqrt{\bar{P}}e^{j\phi{\lbrack n\rbrack}},y\lbrack n\rbrack \geq \rho\sqrt{\bar{P}}} \\
			\end{matrix} \right.
		\end{equation}
where $y[n]$ represents the OFDM time-domain symbolic sample sequence with phase $\phi[n]$, and the phase of the signal does not change before and after limiting. $\rho$ represents the clipping rate (CR), defined as the ratio of the limiting threshold amplitude to the root-mean-square of the original signal symbols.

\subsubsection{$\mu$-law Companding}
The companding method uses a companding function to compress the large-amplitude signal of the OFDM time domain signal and expand the small-amplitude signal, thus reducing the PAPR of the signal. Unlike the clipping method, the companding method is able to recover the signal to some extent by an inverse companding function, and the recovery effect is related to the choice of the companding function.
		
The $\mu$-law companding function is expressed as:
		\begin{equation}\label{eq:7}
			z_{n,k} = \frac{Vy_{n,k}}{\left. {\ln\left( {1 + \mu} \right)}| y \right._{n,k} |}{\ln\left( 1 + \frac{\mu}{V}\left| y \right._{n,k}| \right)}
		\end{equation}
where $y_{n,k}$ represents the $k$-th sample of the $n$-th OFDM symbol of the transmit signal. $V$ denotes the magnitude mean of the time-domain OFDM symbols, and $| y_{n,k} |$ is the modulus of complex samples $y_{n,k}$. $\mu$ is a constant that controls the degree of compression.
	
The inverse companding function applied at the receiver is expressed as:
\begin{equation}\label{eq:8}
{y_{n,k}}^{'} = \frac{V^{'} {z_{n,k}}^{'}}{\mu \left| {z_{n,k}}^{'} \right|} e^{{{(\frac{{|{z_{n,k}}^{'}|} {\ln{({1 + \mu})}}}{V^{'}})} - 1}}
\end{equation}
where $V^{'}$ denotes the average amplitude of the received signal ${z_{n,k}}^{'}$.
		
\subsection{Conventional signal non-distortion techniques}
\subsubsection{SLM}
\begin{figure}[t]
%\vspace{0.18in}
\centering
\includegraphics[width=2.2in]{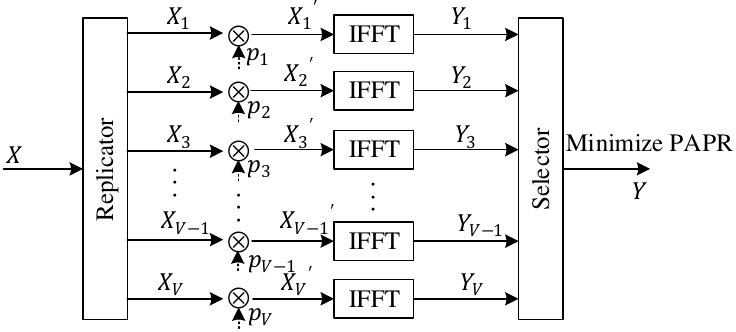}
\caption{SLM system framework.}
			\label{fig_3}
		\end{figure}

The SLM system block diagram is shown in Fig. \ref{fig_3}. The frequency-domain signal $X$ is copied into $V$ copies denoted as $\left\{ X_{1},X_{2},X_{3},\ldots,X_{V - 1},X_{V} \right\}$, and each copy is multiplied by a sequence of phase factors $p_{v},v\in 1,2,\ldots,V$, where a single phase factor ${p_{v}} ^{(k)} = {\mathit{\exp}\left( {j\varphi} \right)},\varphi\in\lbrack 0,2\pi\rbrack$. Then we have the phase-transformed signal set $X^{'}$ and the time-domain OFDM signal set $Y$ by the IFFT transform. The selector selects from the set of signals $Y$ to obtain the smallest PAPR for transmission. The group sequence of the selected signal is transmitted to the receiver as additional information together with the signal, and the receiver can recover the original signal by finding the corresponding phase factor sequence according to the sequence number. In this process, SLM does not bring additional distortion to the original signal.
		
\subsubsection{PTS}
\begin{figure}[t]
%\vspace{0.18in}
\centering
\includegraphics[width=2.5in]{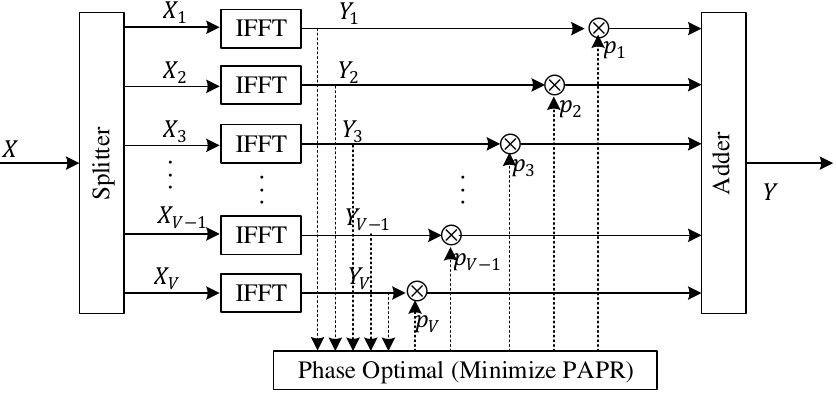}
\caption{PTS system framework.}
\label{fig_4}
\end{figure}

PTS is very similar to SLM and its structure is illustrated in Fig. \ref{fig_4}. A modulated symbol sequence $X = \left\lbrack X_{1},X_{2},X_{3},\ldots,X_{N - 1},X_{N} \right\rbrack$, where $N$ denotes the number of subcarriers, is partitioned into $V$ parts. The number of valid data in each part is $N/V$, and the rest of the positions are filled with zeros to form a data block of length $N$, and then $N$-point IFFT transform is applied to the data block to generate the time domain signal. A random sequence of $V$ phase factors of length $N$ is generated $p_{v},v\in 1,2,\ldots,V$, where a single phase factor ${p_{v}}^{(k)} = {\mathit{\exp}\left( {j\varphi} \right)},\varphi\in\left\lbrack {0,2\pi} \right\rbrack,k\in 1,2,\ldots,N$, usually chosen among $\lbrack \pm 1, \pm 1j\rbrack$. The phase factor sequence and the corresponding signal of each branch are weighted and summed, and the PAPR is calculated. After $M$ times of the same operation, the phase factor sequence with the smallest PAPR is selected as the final phase factor sequence to generate the time-domain emission signal, which would lead to the reduction of PAPR.
		
\subsection{Deep learning-based techniques}
\subsubsection{PAPR Loss}
According to \cite{51}, adding PAPR to the loss function of a neural network can learn a network model with low PAPR output. To reduce the distortion between the output and input of the network, the loss function $L_{1}$ of the original DJSCC can be expressed as:
\begin{equation}\label{eq:04:14}
L_{1} = \left| \left| {\boldsymbol{x} - g\left( {FFT\left( IFFT\left( {f\left( {\boldsymbol{x},\theta} \right)} \right) + \boldsymbol{n} \right),\phi} \right)} \right| \right|_{2}
\end{equation}
To reduce the PAPR, the loss function $L_{2}$ can be expressed as:\begin{equation}\label{eq:04:14}
L_{2} = \mathbb{E}\left\lbrack {PAPR\left\{ {IFFT\left( {f\left( {\boldsymbol{x},\theta} \right)} \right)} \right\}} \right\rbrack
\end{equation}
where $\mathbb{E}$ denotes the mean value.

%        \addtolength{\topmargin}{0.17in}
In order to minimize both distortion and PAPR, the new DJSCC loss function is constructed by combining the loss functions $L_{1}$ and $L_{2}$ denoted as:
\begin{equation}\label{eq:04:14}
L = L_{1} + \lambda L_{2}
\end{equation}
where $\lambda$ denotes the weight to trade-off the two penalties. A larger $\lambda$ means that the learning is more biased towards reducing the PAPR and vice versa for reducing the distortion between the reconstructed signal and the original signal.

%        \addtolength{\topmargin}{0.15in}

\subsubsection{Clipping with retraining}	
Deep learning can learn the output of a network with certain features through a nonlinear function in the network. Therefore, clipping can be added as a nonlinear function to the network in the training process to further improve the performance. The clipping function can be expressed as follows:
\begin{equation}\label{eq:04:15}
y_{clip}[n] = y[n] \left( {1 - \frac{ReLU\left( \left| y[n] \right| - \rho \overset{-}{y}[n] \right)}{\left| y[n] \right| + \gamma}} \right)
\end{equation}
where $y[n]$ represents the IFFT transformed signal of the joint encoder output, $\rho$ represents the clipping ratio, $\left| {y[n]} \right|$ represents the mode of the complex symbol, $\overset{-}{y}[n]$ represents the average amplitude of the signal, and $\gamma$ represents a constant much smaller than 0 to avoid the denominator being 0.
		
\begin{figure}[t]
%\vspace{0.2in}
\centering
\includegraphics[width=1.73in]{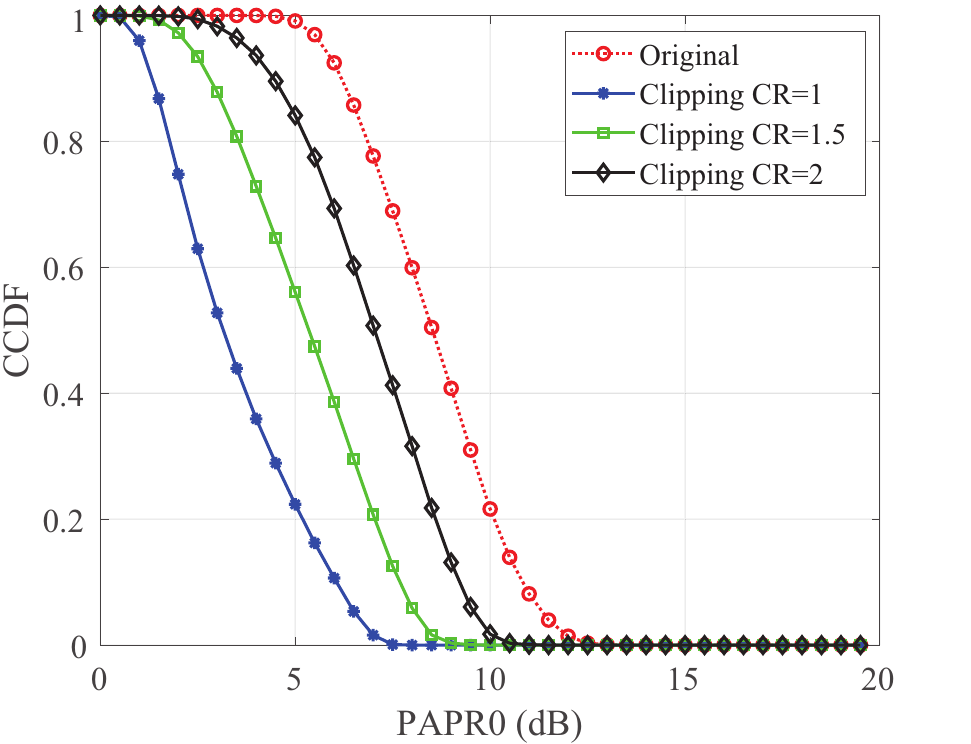}
\includegraphics[width=1.7in]{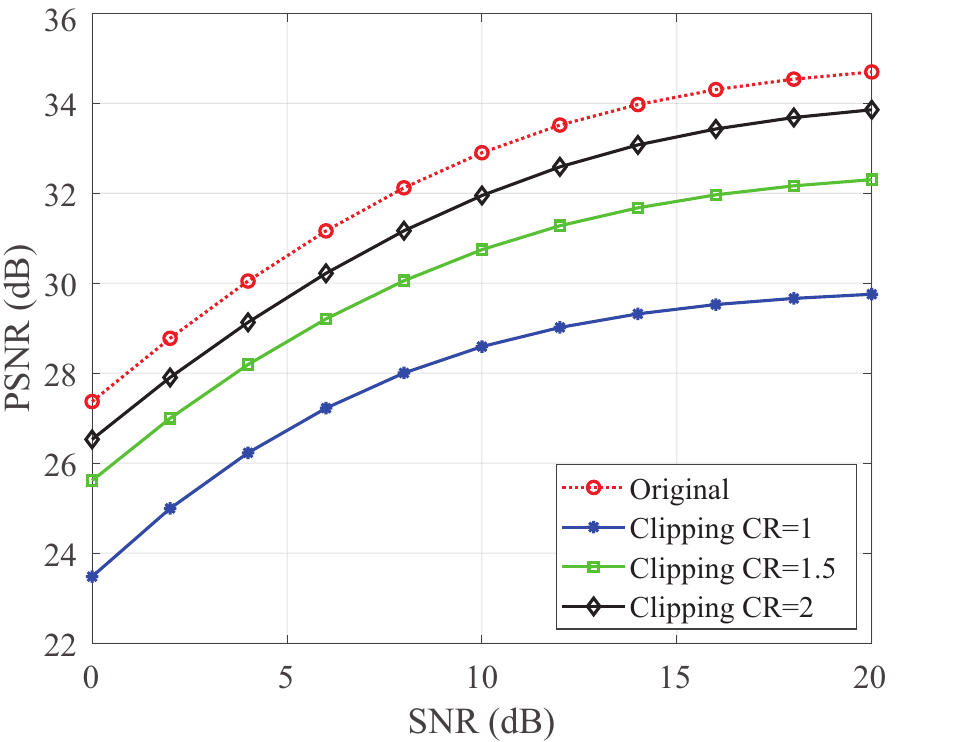}
\caption{Performance of PAPR (left) and PSNR (right) of clipping in DJSCC.}
\label{fig_5}
		\end{figure}
		
		\begin{figure}[t]
			
			\centering
			\includegraphics[width=1.73in]{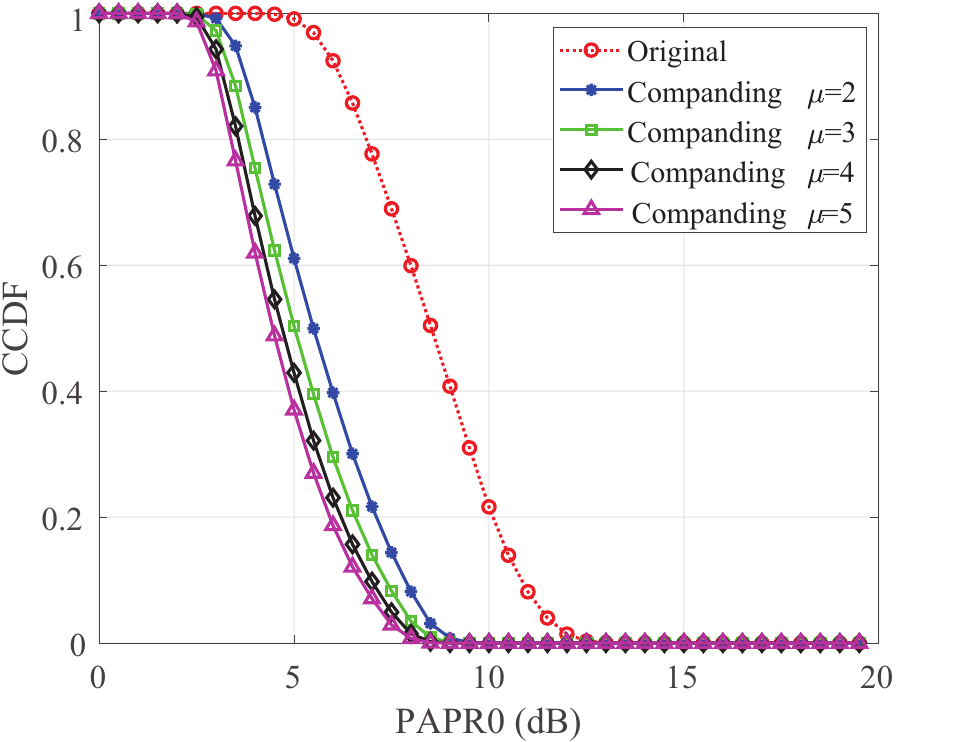}
			\includegraphics[width=1.7in]{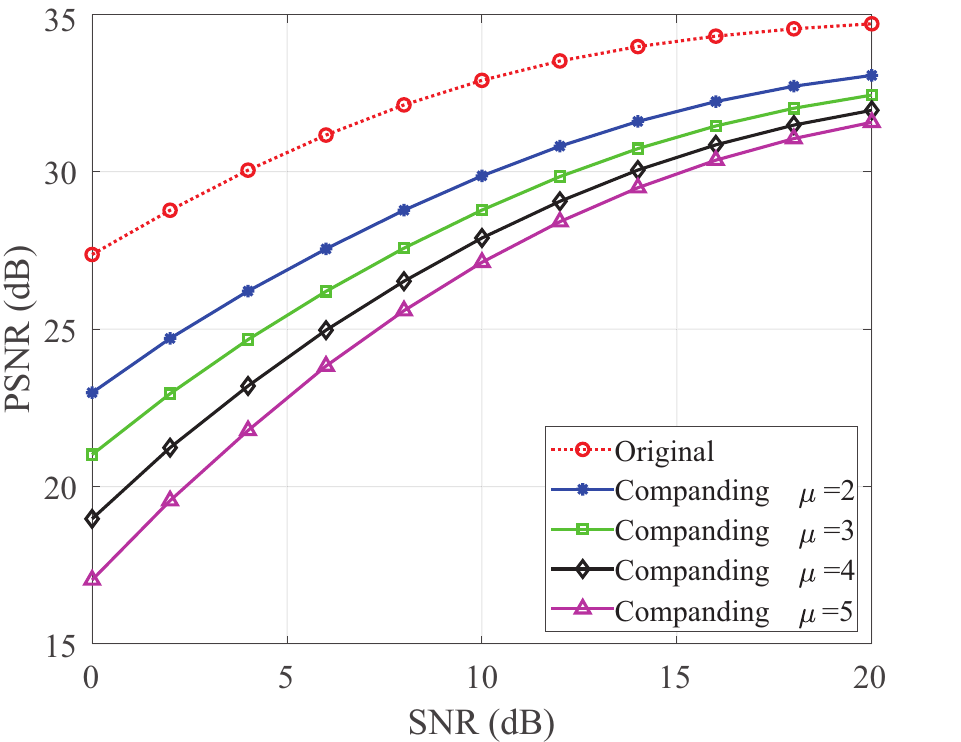}
			\caption{Performance of PAPR (left) and PSNR (right) of companding in DJSCC.}
			\label{fig_6}
		\end{figure}
						 	
\section{Experimental Results}
We used tensorflow and its high-level API Keras to learn the encoder and decoder of DJSCC, and an Adam optimizer with a learning rate of $10^{-4}$ and a batch size of 128 was considered in the training. ImageNet \cite{ref22} was used as the training dataset, which contains more than 14 million images with more than 21,000 categories. We train the model with the SNR drawn from a uniform distribution in the range of $[0,20]$ dB. For simplicity, the neural network is trained under the AWGN channel, and extension to the fading channel is trivial.
%		since slow fading channel can be approximated as AWGN channel after the equalization process.
The Kodak dataset was used to evaluate DJSCC performance, which contains 24 images of size $768 \times 512$. Each image in the Kodak dataset is transmitted 10 times with different noise instances. In all experiments, the number of OFDM subcarriers is 64, and the DJSCC bandwidth to signal ratio is set to 1/6.
		
\begin{figure}[t]
%\vspace{0.1in}
\centering
\includegraphics[width=1.72in]{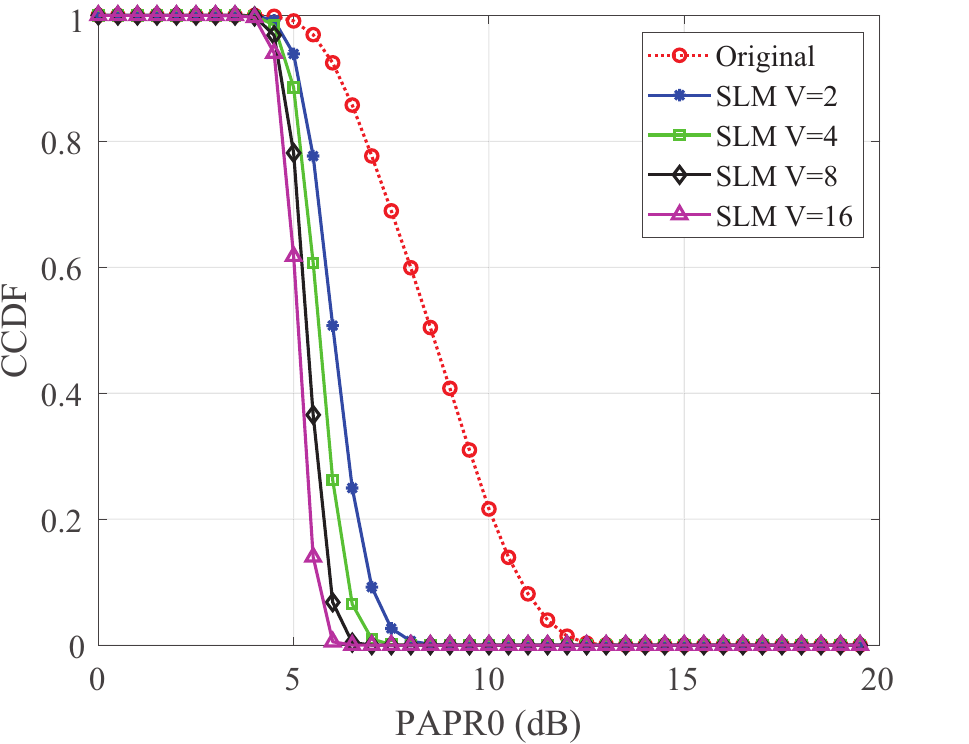}
\includegraphics[width=1.72in]{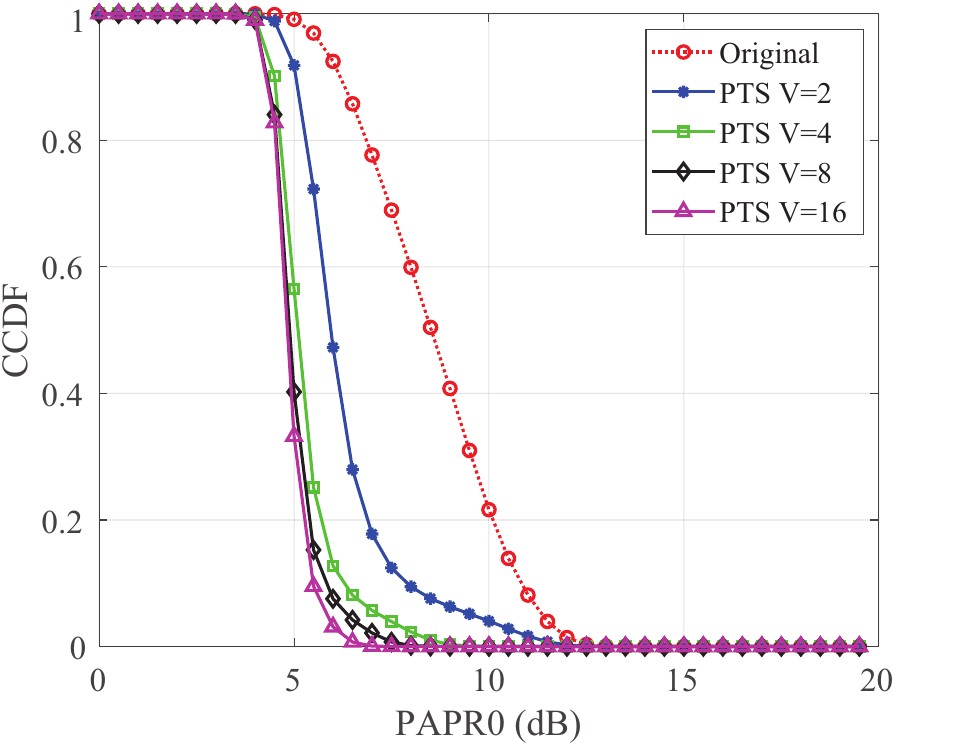}
\caption{Performance of PAPR  of SLM (left) and PTS (right) in DJSCC.}
\label{fig_7}
\end{figure}

Fig. \ref{fig_5} shows the PAPR and image peak signal to noise ratio (PSNR) performance of DJSCC after conventional clipping. It can be seen that the PAPR performance of DJSCC has improved significantly after clipping, and the improvement increases as the CR decreases. However, the signal distortion generated by clipping also leads to the degradation of PSNR performance, and the magnitude of the degradation increases with the decrease in CR. The performance of DJSCC using the companding method is shown in Fig. \ref{fig_6}. It can be seen that the PAPR performance of DJSCC after companding has improved significantly, but the PSNR performance has decreased. Furthermore, the improvement in PAPR increases with increasing $\mu$, while the quality of image reconstruction decreases with increasing $\mu$. We assume that the phase factor sequences of SLM and PTS are transmitted without errors, then they do not cause distortion to the signal, so we only observe the PAPR reduction performance of both as shown in Fig. \ref{fig_7}. It can be seen that both methods can inhibit PAPR in DJSCC, and the suppression level increases with the increase of $V$.
        	
		     \begin{figure}[t]
		\centering
		\includegraphics[width=1.73in]{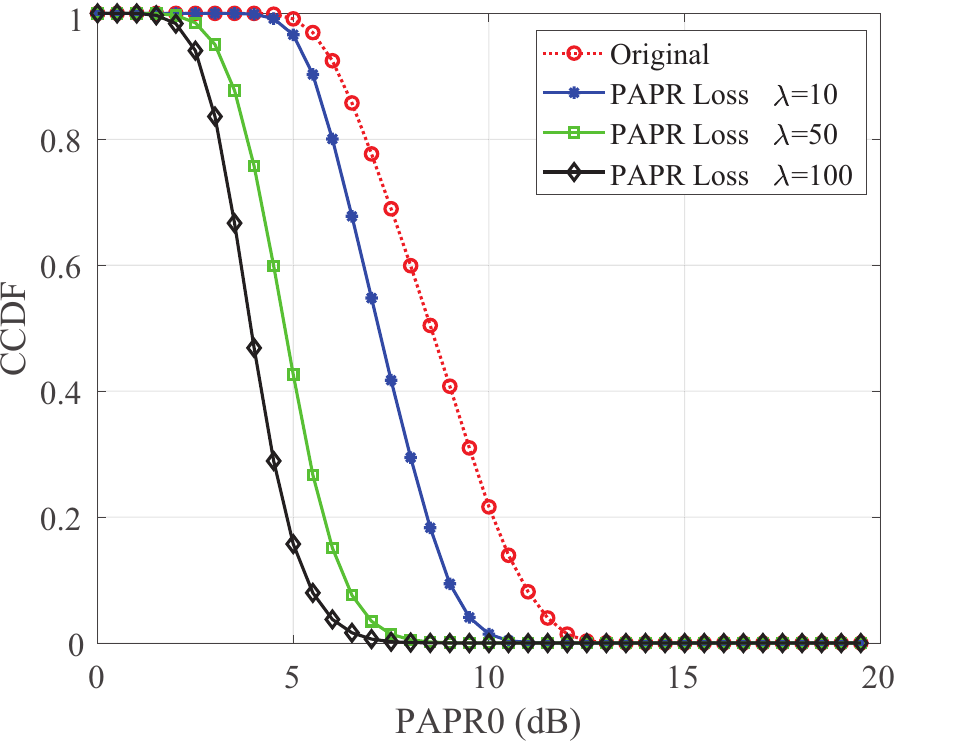}
		\includegraphics[width=1.7in]{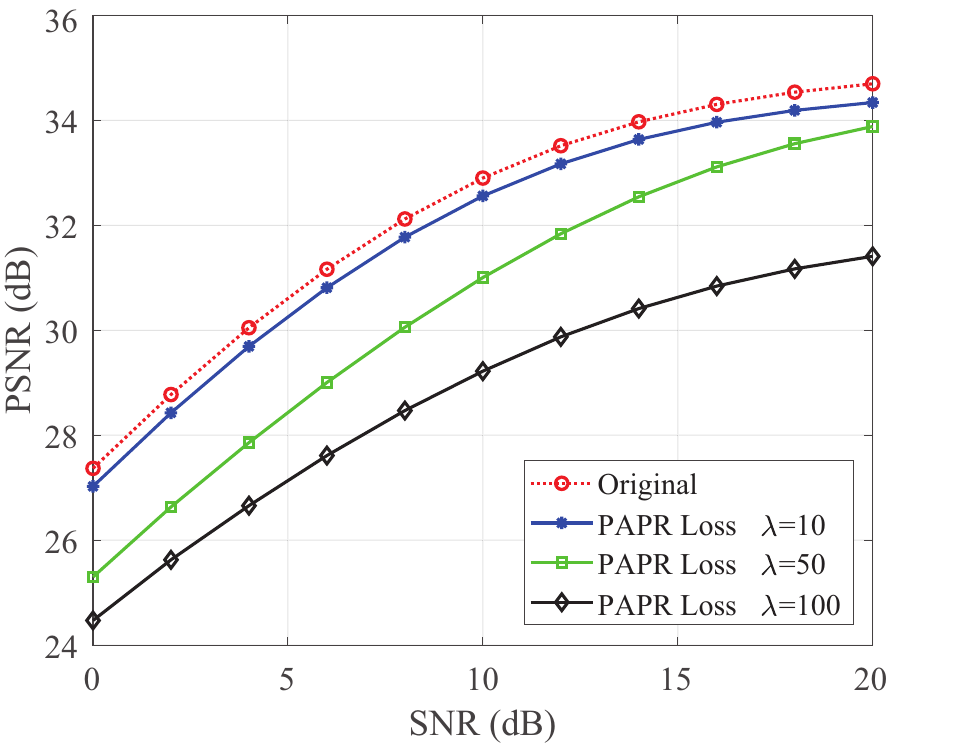}
		\caption{Performance of PAPR (left) and PSNR (right) of ``PAPR Loss" in DJSCC.}
		\label{fig_8}
	\end{figure}	
	\begin{figure}[t]
		%\vspace{0.1in}
		\centering
		\includegraphics[width=1.73in]{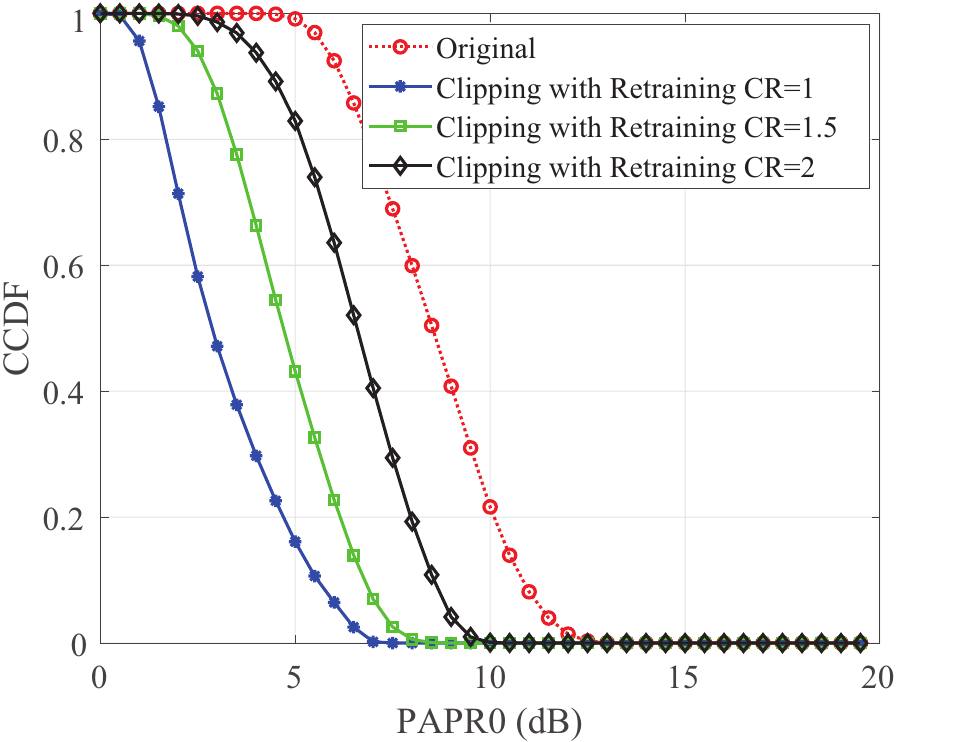}
		\includegraphics[width=1.7in]{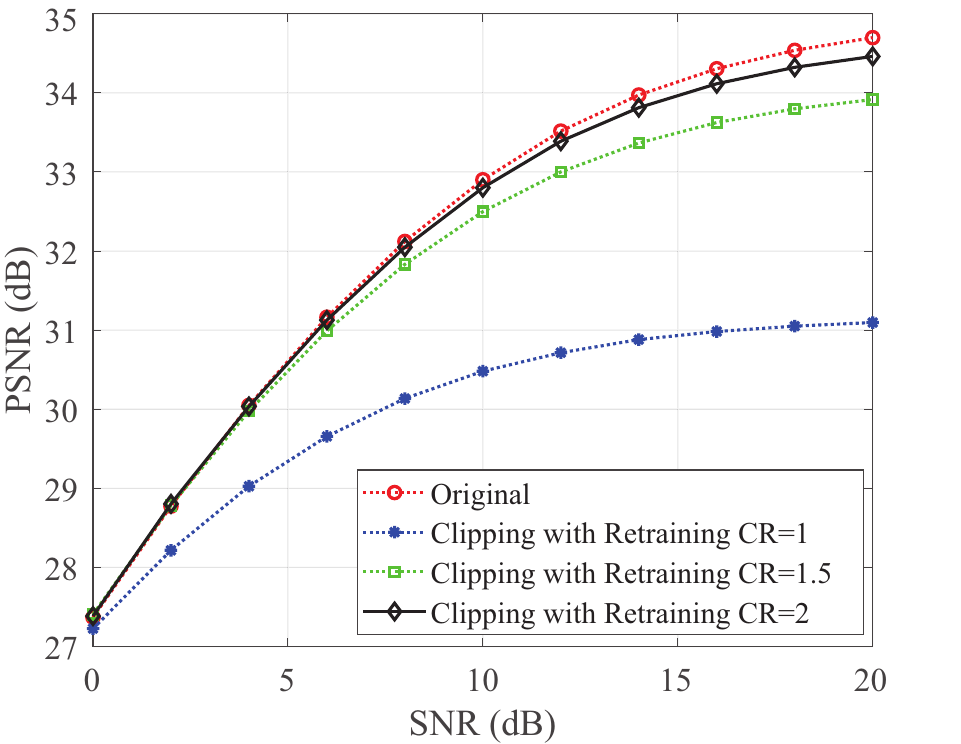}
		\caption{Performance of PAPR (left) and PSNR (right) of ``Clipping with Retraining" in DJSCC.}
		\label{fig_9}
	\end{figure}	

Fig. \ref{fig_8} shows the performance of DJSCC after adding PAPR to the loss function. It can be seen that this method can suppress the PAPR of DJSCC, but it also reduces PSNR performance. Moreover, both the PAPR of DJSCC and the reconstruction quality of the images decrease with increasing penalty factor $\lambda$. Fig. \ref{fig_9} demonstrates the PAPR and PSNR performance by adding the clipping process to the DJSCC training. It can be seen that this method reduces the PAPR of DJSCC while degrading the PSNR performance. The decrease in PSNR is smaller compared to the conventional clipping method with the same CR, especially when the CR is larger.

\begin{figure}[t]
		    \centering
\includegraphics[width=1.73in]{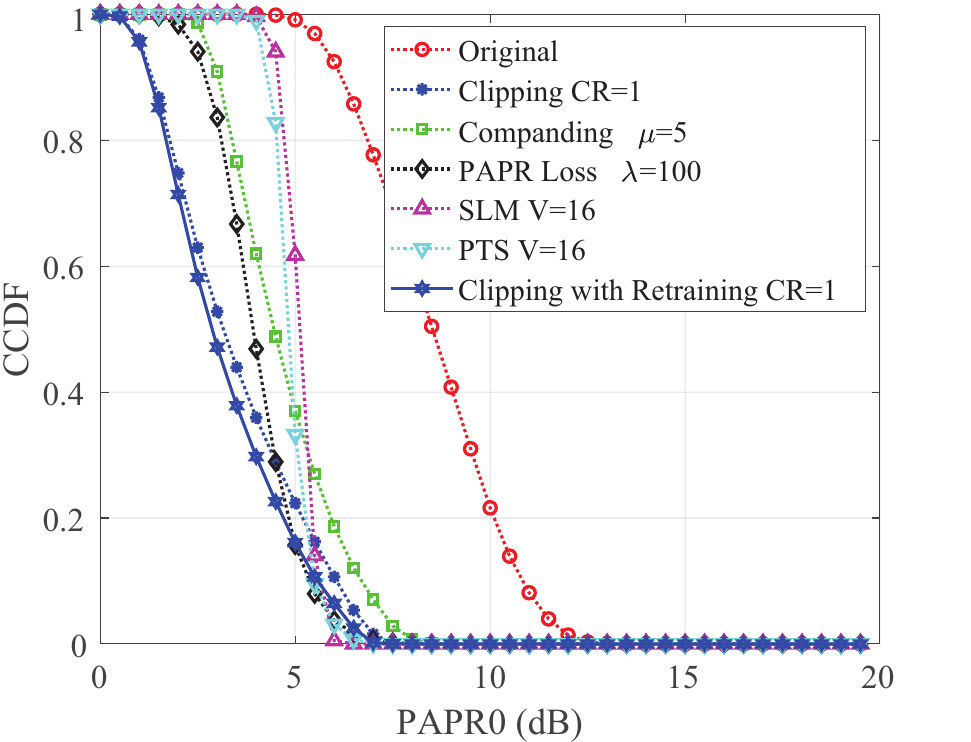}
\includegraphics[width=1.7in]{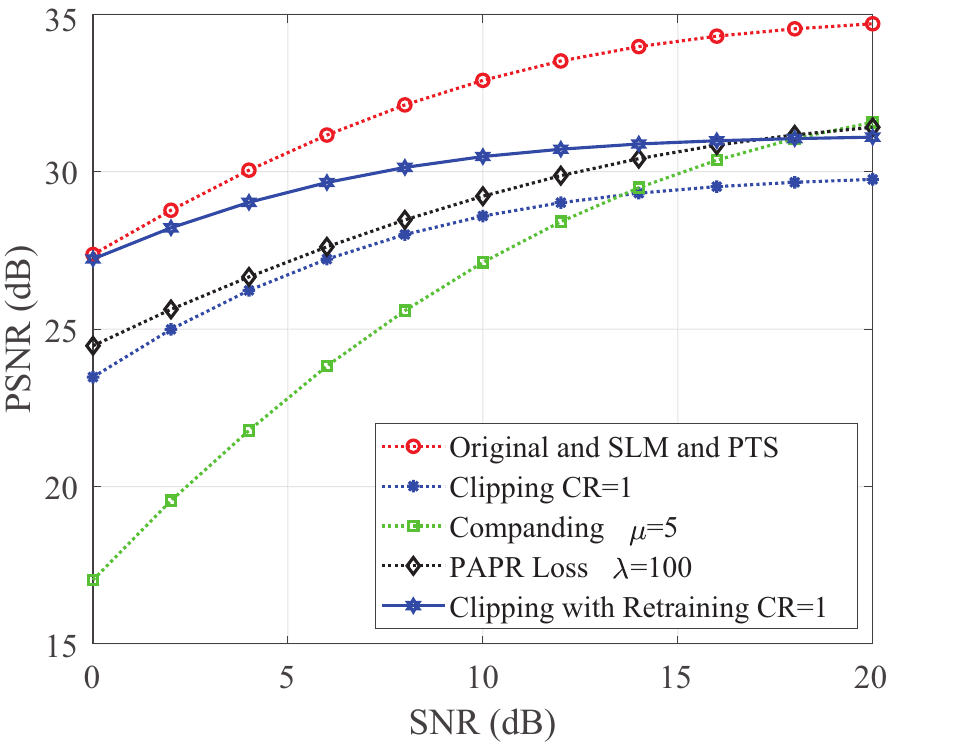}
\caption{Performance comparison of all PAPR reduction techniques in DJSCC.}
		    \label{fig_10}
        \end{figure}

We evaluate the PAPR and PSNR performance of all the methods mentioned in the previous section, shown in Fig. \ref{fig_10}. Among signal distortion PAPR reduction techniques, the clipping with retraining achieves the best PSNR performance, especially at low SNRs.
% For example, at SNR=0, the recovered image at the receiver is about 3 dB higher than that of the PAPR penalized loss function method and the conventional clipping, and about 10 dB higher than that of the companding.
However, in conventional SSCC with digital modulation, the performance of the companding method is superior to that of the clipping method, especially at higher order modulations\cite{67}. In addition, clipping with retraining does not require any modification of the network model and does not add excessive computational complexity. Furthermore, we also find that signal non-distortion PAPR reduction techniques, i.e., SLM and PTS, can successfully reduce the PAPR in DJSCC without compromise to signal reconstruction.

\section{Conclusion}
In this paper, we investigate PAPR reduction techniques for DJSCC-based OFDM system, including conventional PAPR reduction techniques and DL-based PAPR reduction techniques. Simulation results show that although the conventional PAPR reduction techniques can be applied to DJSCC, the performance in DJSCC is different to that of conventional SSCC, e.g., the performance of clipping is better/worse than companding in DJSCC/SSCC. Moreover, for signal distortion PAPR reduction techniques, clipping with retraining achieves the best performance in terms of both PAPR reduction and recovery accuracy. Furthermore, signal non-distortion PAPR reduction techniques, i.e., SLM and PTS can successfully reduce the PAPR in DJSCC without compromise to signal reconstruction.

		%	\begin{thebibliography}{00}
			%		\bibitem{ref1}
			%		C. E. Shannon, "A mathematical theory of communication," in The Bell System Technical Journal, vol. 27, no. 3, pp. 379-423, July 1948.
			%		
			%		\bibitem{ref2}
			%		E. Bourtsoulatze, D. Burth Kurka and D. Gündüz, "Deep Joint Source-Channel Coding for Wireless Image Transmission," in IEEE Transactions on Cognitive Communications and Networking, vol. 5, no. 3, pp. 567-579, Sept. 2019, doi: 10.1109/TCCN.2019.2919300.
			%		
			%		\bibitem{ref3}
			%		D. B. Kurka and D. Gündüz, "Successive Refinement of Images with Deep Joint Source-Channel Coding," 2019 IEEE 20th International Workshop on Signal Processing Advances in Wireless Communications (SPAWC), 2019, pp. 1-5, doi: 10.1109/SPAWC.2019.8815416.
			%		
			%		\bibitem{ref4}
			%		D. B. Kurka and D. Gündüz, "DeepJSCC-f: Deep Joint Source-Channel Coding of Images With Feedback," in IEEE Journal on Selected Areas in Information Theory, vol. 1, no. 1, pp. 178-193, May 2020, doi: 10.1109/JSAIT.2020.2987203.
			%		
			%		\bibitem{ref5}
			%		D. B. Kurka and D. Gündüz, "DeepJSCC-f: Deep Joint Source-Channel Coding of Images With Feedback," in IEEE Journal on Selected Areas in Information Theory, vol. 1, no. 1, pp. 178-193, May 2020, doi: 10.1109/JSAIT.2020.2987203.
			%		
			%		\bibitem{ref6}
			%		Schmidl, Timothy M., and Donald C. Cox. "Robust frequency and timing synchronization for OFDM." IEEE transactions on communications 45.12 (1997): 1613-1621.
			%	\end{thebibliography}
		\bibliographystyle{ieeetr}
		\bibliography{ref.bib}
		
		\vspace{12pt}
		%    \end{spacing}	
\end{document}